\documentclass[prd,nobibnotes,nofootinbib,showpacs,12pt]{revtex4}
\pdfoutput=1
\usepackage[latin1]{inputenc}
\usepackage{graphicx}
\usepackage{amsmath}

\numberwithin{equation}{section}
\usepackage[mmddyyyy]{datetime}
\newdate{date}{20}{09}{2020}

\begin{document}
\title{The quenched Eguchi-Kawai model revisited.}
\author{Herbert Neuberger}
\email{herbert.neuberger@gmail.com}
\affiliation{Department of Physics and Astronomy, Rutgers University,\\ 
Piscataway, NJ 08854, U.S.A} 

\begin{abstract}
The motivation and construction of the original Quenched Eguchi-Kawai model are reviewed,
providing much greater detail than in the first, 1982 QEK paper. 
A 2008 article announced that QEK fails as a reduced model because 
the average over permutations of eigenvalues stays annealed. It is shown here that the original quenching logic naturally leads to a formulation with no annealed average over permutations. 

\end{abstract}
\date{\displaydate{date}}
\pacs{11.15.Ha, 11.15.PG
}
\maketitle

\section{Introduction.}

In 1982 Eguchi and Kawai (EK)~\cite{ek} made the spectacular observation 
that $SU(N)$ lattice gauge theory
with no matter fields gave Wilson loop expectation values that were
reproducible on a minimal $1^4$ lattice  at leading order in $1/N$.   
Intensive research ensued. It was found that large $N$ phase transitions
caused difficulties in approaching the continuum limit. Some cures
were suggested. The purpose of this paper is to make explicit the underlying
logic and definitions of one of them: the QEK conjecture~\cite{qek}.  
It will be shown that the 2008 article~\cite{bs} does not establish failure of
the general idea of quenching. A determination of the validity or failure of 
quenched reduction requires future extensive numerical work.

\section{Eguchi-Kawai reduction is not just a lattice peculiarity.}

At first sight EK reduction seemed to be an incredible lattice trick.
Only much more recently has numerical work indicated that EK reduction
is a property of {\underline{continuum}} planar four dimensional $SU(N)$ pure gauge
theory~\cite{contek}. At leading order in $1/N$ the expectation values of Wilson 
loops on an infinite Euclidean four-torus are the same as ``folded'' 
counterparts on any finite torus with side size larger than about one Fermi in 
QCD terms.

In retrospect, this sharpens the original question that motivated~\cite{qek}: 
how is a Coulomb-like law realized in a finite periodic Euclidean four volume 
which has no room for ``Faraday's flux lines'' to spread? 

Continuum EK reduction is some times referred to as ``partial reduction''
because the lattices one has to use in a numerical simulation in order 
to approach the continuum limit to reasonable accuracy 
must be substantially larger than $1^4$. ``Partial'' 
is a bad epithet because it emphasizes a detail of implementation and conceals 
the physical content of the result. 

\section{Lattice fix of EK reduction.}

The first conjectured fix for lattice EK reduction to an $1^4$ lattice was the quenched EK model, QEK~\cite{qek}.

The main physics question was how the largeness of the $SU(N)$ group at infinite $N$ could provide a place-holder for an infinite lattice while consisting of just four unitary link matrices. A simple calculation at one loop order showed
that the eigenvalue phases of the link matrices in the four directions could play the role
of continuous lattice momenta in $(-\pi, \pi]^4$ -- an ``emerging'' toroidal momentum space -- and produce the standard Coulomb's force law on the lattice {\underline{if}} we 
summed up the contributions of a large number of saddles and ignored their instabilities.
As was very well understood from other semi-classical calculations, 
fluctuations in the ``flat'' directions, connecting the saddles, produced zero modes and were easily dealt with.
But, the saddles in the integral were dominated by coalescing eigenvalues, so perturbation theory was unstable. As a whole, the matrix integral was benign. 

On an $1^4$ lattice the eigenvalue sets of each link matrix are gauge invariant angles. They
are unique candidates for lattice momenta. 

The QEK fix consisted of a removal of the link matrices' eigenvalues from the set of annealed
variables. They were quenched instead. This difference did not matter at large $N$ 
by a degrees-of-freedom counting argument: There were $4N$ angles, and order $4N^2$ 
matrix elements: The angles ought to be governed by uniform and uncorrelated distributions
in each direction if some obvious symmetries remain preserved as $N\to\infty$. 

For strong lattice coupling (small ``$\beta$''), where EK had been proven to work,  
the quenching prescription would have no effect to leading order in $\frac{1}{N}$. Quenched or annealed, the angle distributions would be frozen to continuous uniform densities in each direction and those would be uncorrelated. The requirement to match onto the original EK version, which was {\underline{proven}} to hold at strong coupling, left little freedom for constructing QEK. The ``loop equations'', on which the EK proof relied, have trivial boundary conditions in the strong coupling limit and determine the entire strong coupling series. That 
series has a finite radius of convergence. The precise boundary conditions for the lattice loop equations at weak coupling remain unknown to date. They would be needed for constructing the Feynman series for Wilson loops. The loop equations themselves have only a relatively formal continuum limit. They do not offer a reliable tool for analysis in continuum directly. 

\section{Quenching in detail}

QEK was originally presented as a conjecture and 
this remains its status to date. It is uncertain whether it is valid 
throughout the bridge connecting short and long distance pure gauge theory physics. 
Even if it does, there remains doubt whether 
it would be practical in comparison to the safer 
continuum EK method which relies both on the lattice loop equations and
on some numerical, nonperturbative tests. 

The QEK prescription is explained below in detail and at an elementary level. 

\subsection{Quenching ``with calculus''}

The lattice variables we shall deal with are phase angles and unitary matrices.

The first step in constructing the quenching prescription 
consists of a {\underline {proper}} change of integration variables in the EK case. For
simplicity, we consider $U(N)$ -- restricting to $SU(N)$ later presents little difficulty. The change of variables
requires a one-to-one relation between the old and new, together with a matching of integration domains. 
The domains in the new variables are obviously important and will be discussed later below.

The variable change is an EK$\to$QEK map replacing each of the link variables 
$U_\mu , \mu=1,2,3,4$ by a set of angles
$\theta_\mu^i$, $i=1,...,N$ and a unitary matrix $V_\mu$. 
For four fixed $U_\mu$'s , there are multiple solutions. To determine
domains of integration in the new variables requires selecting
one unique branch among them. 

\begin{equation} (1)~~U_\mu V_\mu = V_\mu D_\mu (\theta_\mu );~~~~~~~~~(2)~~U_\mu  = V_\mu D_\mu (\theta_\mu )V_\mu^\dagger.
\label{UtoangV}\end{equation}
where
\begin{equation}
D_\mu (\theta_\mu ) = {\rm diag}(e^{i\theta_\mu^1}, e^{i\theta_\mu^2},....e^{i\theta_\mu^N} ). 
\end{equation}

The original integration measure, is 
\begin{equation}
\prod_{\mu=1,2,3,4} dU_\mu {\rm ~~ {where~~}}dU_\mu {\rm ~~{is~Haar.}}\end{equation}

After the change of variables the integration measure is {\underline{locally}} given by 
\begin{equation}
\prod_{\mu=1,2,3,4} [ dV_\mu \prod_{i=1,...N} d\theta_\mu^i \prod_{i<j} \sin^2 (\frac{\theta^i_\mu - \theta^j_\mu}{2})  ] \label{jac},\end{equation}
up to a constant factor determined by a normalization convention. 
The Jacobian does not depend on $V$, only on the angles. This is crucial for the quenching proposal because the integration over the angles factorizes. In EK the instability resides in the the angle dynamics. The QEK model addresses the instability directly.

The $U_\mu$'s are uniquely determined by the $\theta$'s and $V$'s: 
The $\theta_\mu$ exponents are the eigenvalues of $U_\mu$
and the columns of $V_\mu$ are the corresponding eigenvectors. The multiple solutions
of equality (2) in eq.~\ref{UtoangV} with given $U$ are all generated from each other 
by group transformations. This makes the search for unique branches straightforward. 
Two types of maps between solutions enter and 
together they exhaust their multiplicity:

First, one may multiply equality (1) in 
eq. \ref{UtoangV} by a $\mu$-dependent diagonal matrix from the {\underline{right}}. 
This multiplies each column of $V_\mu$ by a phase: eigenvectors are rays. 
The transformations are a product of commuting $U(1)^N$'s, one for each direction. 
Second, one may multiply equality (1) in 
eq. \ref{UtoangV} from the {\underline{right}} by a permutation 
matrix $P\in S(N)\subset U(N)$.
$P$ is a unitary matrix having a single entry equal to 1 in each row and each column. 
When acting on a column from the {\underline{left}},   
it permutes its row entries by $P$. $P^\dagger$ acts from the {\underline{right}} on a row and 
permutes its column entries by $P^{-1}$. 
Inserting $1=P P^\dagger $ on the
right hand side of of eq. (1) in \ref{UtoangV} between the V and the $\theta$-diagonal matrix replaces 
$V$ by $VP$ on both sides. The effect of $P^\dagger ..... P$ on the diagonal $\theta$ matrix
is to permute its columns by $P$ and its rows by $P^\dagger$. The result is that
the $\theta_\mu^i$ get permuted by $P$ along the diagonal. Thus, elements of $P\in S(N)$ act simultaneously on the columns of $V$ and on their associated eigenvalues, preserving the  eigenvalue--eigenvector association. 

Suppose one is given a $U$ matrix. By a probability argument, it has distinct eigenvalues. 
A procedure to identify a unique decomposition in terms
of $V$ and $\rm diag(\theta)$ matrices is defined as follows: One first finds all the roots of 
the characteristic polynomial of $U$. Then, for each eigenvalue, one finds
a corresponding unit norm column eigenvector. Arranging the columns left to right, one gets
a unitary matrix $V$. This $V$ has its columns further reordered into a canonical form. 
This form will be defined below. The positions of the angles
along diagonals of the $D$ matrices are fixed by the order of the columns of the matrix $V$.

We learned that the domain of integration in $V$ and $\theta$ variables is the $S(N)$  
right coset space ~\cite{eynard}:
\begin{equation}
[ U(N)/U(1)^N \times U(1)^N ] /S(N) \label{coset}.
\end{equation}
Factoring by $S(N)$ defines an equivalence relation $\sim$ in the space of the new variables which 
partitions $U(N)/U(1)^N \times U(1)^N$ into equivalence classes. The domain of integration is chosen to be one 
specific representative of each class. The intention is to make a choice that has a chance
to protect the perturbatively emerging momentum space from the instability: We choose in each class a representative element of the
matrix component, $V$, close to the unit
matrix. This element consists of an ordered set of column eigenvectors of the corresponding $U$-matrix.
It can also be viewed as a choice of an ordered orthonormal basis whose elements correspond to a set of $N$ angles determining the eigenvalues of the corresponding $U$ matrix. The order
of the eigenvalues is determined by the order of the columns of $V$. The solution for $V$ and $\theta$ of equality (2) in~\ref{UtoangV} has become unique.  

The prescription suggested here is that $V$ have the entry of maximal 
absolute value along the portion of each row from the diagonal to the right on the diagonal. Equal absolute values are ignored for probability reasons. That entry is 
made real positive by a phase choice for the corresponding column.
A given $V$ can be brought into this configuration sequentially, starting from the
top row and bringing the desired entry to the (1,1) location by permuting through $N$ columns, and then continuing 
to the next row and permuting through the $N-1$ remaining columns to bring the desired element
of the row to the $(2,2)$ location and so on. Real positivity fixes the phase of each column.
It mods out the factor $U(1)^N$ in the ``numerator'' of ~\ref{coset}.
For the lack of a better term, let us call this ordering ``diagonally right row entry dominant''. This clumsy name avoids confusion with the standard term ``diagonal dominance''.
With probability one, this choice is unambiguous.

In the EK case, at any finite $N$, which specific choice of representative was made does not matter because the integral over 
ordered angles and bases is done on one common integrand and the $U(1)^N$ and $S(N)$ 
are symmetries of the integrand.
In the QEK case however, the integral over the angles occurs at
a later stage, with a modified integrand and the choice does matter because 
$S(N)$ acts simultaneously on $\theta$'s and $V$'s. At finite $N$ quenched expectation values 
depend on which representative of each equivalence class was chosen. This is a crucial 
feature of quenching.

The correct ranges of integration in QEK are over  a set of four $V$'s, all in ``canonical'' order. A criterion for choosing the canonical order is that the $V$ representative 
be ``perturbative''. There are no restrictions on the angles. Different orders of the
same set are included as separate contribution. 
There may exist other prescriptions that are equally valid. To be sure, 
for large lattice $\beta$ couplings the reduction validity of QEK remains a conjecture. 

Permutations of eigenvectors must be eliminated in order to perform a correct variable change in EK with no over-counting at any coupling. That is just applied multivariate calculus.

To motivate the ordering prescription 
for the $V$ matrices, consider the simple case of $U(2)/S(2)$. The two $U(1)^2$ terms in the ``numerator'' of eq.~\ref{coset} are ignored.  
Up to an irrelevant overall phase, any $V\in U(2)$ can be written as
 \begin{equation}
 V=
 \begin{pmatrix}
 w & z^*\\
 -z & w^*
 \end{pmatrix}.
 \end{equation}
 where $|w|^2+|z|^2=1$. $S(2)=\{ 1_2 , \sigma_1\}$ using Pauli's notation. Let $V^\prime$ be given 
 by the same expression with $w,z$ replaced by $w^\prime, z^\prime$. The 
 equation $V=V^\prime \sigma_1$ defines an equivalent pair $V^\prime \sim V$. In components, 
 $z^\prime = -w^*$ and $w^\prime = - z^*$. For $V$ to be  ``diagonally right row entry dominant''
 we need $|w|>|z|$; then $V^\prime$ is not ``diagonally right row entry dominant'' 
 because $|w^\prime| <|z^\prime |$. The split of $U(2)$ 
 into ``halves'' is explicit. The half containing the identity makes up the QEK integration domain for $V$. As usual, ambiguous cases are ignored for probability reasons.

 $U(N)$ can be restricted to $SU(N)$ by adding a factor of $\delta_{2\pi} (\sum_i \theta_\mu^i)$ in the $\theta$ integral
 for each direction. $\delta_{2\pi}$ is the $2\pi$-periodic $\delta$ function.
 Mentally, one can imagine the $V$'s to be also fixed by an overall phase, restricting to $SU(N)$; this phase does not enter observables dependent only on the $U$'s.

Once the change of variables in the EK integrand is correctly implemented one can replace each $U_\mu$ 
by equality (2) in eq. ~\ref{UtoangV} in the EK model and nothing has changed. The integral for the partition function 
can be done successively, first integrating over all the columns of the $V_\mu$ in the pair [$\theta^i_\mu$~~,~~ $i$-th column of $V_\mu$]
with local Haar measure at fixed, ordered, angles at each $\mu$. Next, one integrates over all of the possible  ordered 
angle sets for each $\mu$. 

Quenching replaces the EK by QEK. In QEK the integral for the partition function is replaced by an integral with the same 
measure and action, but at fixed angles in all
directions. The QEK partition function is a function of these angles, $Z(\theta)$. For a Wilson loop observable one uses 
$Z(\theta)$ as normalization, now in the denominator, obtaining averages of the observable at a fixed ordered $\theta$ set. Next, these annealed $V$-averages are integrated over the angles
with weight given by the Jacobian in eq. ~\ref{jac}. The $\theta$ variables are treated as a set of random couplings, akin to the $J$-couplings of a spin-glass model.

The traditional choice for angle  ordering is descending along the diagonal with values in the
segment $(-\pi, \pi]$. Such an ordering makes the distribution equal to the derivative of a smooth 
approximation to the angle dependence on the index in each direction separately~\cite{gw}.
But, this is not permitted in QEK. The order of the angles cannot be restricted in any way.
One has a diagonalizing ordered basis and one can assign to each eigenvector an eigenvalue on the unit circle, distributed just according to the Jacobian factor. The Jacobian measure 
is invariant under direction dependent permutations. They  are not
induced by annealed generation of permutations among the columns of $V$ because 
two distinct ``diagonally right row entry dominant'' $V$-matrices cannot be related by
a nontrivial permutation. 

There is no invariance under the hypercubic group at fixed $\theta$. 
It could happen that new large $N$ transitions occur, to phases where
the hypercubic symmetry is spontaneously broken. 
Such phases indeed do occur in the EK model~\cite{contek}. 
If they persist to the quenched case, QEK fails. 
It also could happen that it is practically impossible to attain high enough
values of $N$ because prohibitively large samples of the eigenvalue sets are needed for a reasonable accurate estimate of the final angle integral.

The ordering of $V$ determines that of $V^\dagger$. $V^\dagger$ is not 
``diagonally right row entry dominant''. 
The action controlling the $V$ average depends only on the six combinations $V_{\mu\nu}=V_{\nu\mu}^\dagger \equiv V_\mu^\dagger V_\nu$ for $\mu > \nu$~\cite{qek}. 
These are overlap matrices of ordered eigenvector-sets corresponding to the angles 
in the $\mu$, $\nu$ directions. The common canonical ordering of the $V_{\mu}$'s  
induces some preference for the $V_{\mu\nu}$'s to be closer to identity with no direct feedback on the angles -- unlike in the EK situation. The hope is that angles are now free to
take on the role of an ``emergent'' lattice momentum space.

\section{``Gauge invariance'' in QEK.}

On an $1^4$ lattice, gauge theory 
has a symmetry under simultaneous conjugation by the same matrix
of all link matrices. These matrices can be thought of as Polyakov loops. Evidently, 
on a one site lattice there can be no geometrically open contours. 

This symmetry acts on the $V$-matrices from the {\underline{left}} and therefore
 commutes with the action on $V$ by the $S(N)$ we had to mod out by. A permutation gauge transformation
 will permute the rows of each $V$. After its action each $V$ needs to be reordered back 
 to canonical order. The end result is that gauge transformations which happen to be
 permutations do not change anything. We might as well forget about them altogether. 

\subsection{Conclusion of section.}

This paper contains the full description of the original, with no shortcuts allowed, QEK model. 
There are many ways and points of view in which QEK can fail.
To the limited extent I understand it, the Bringolz-Sharpe paper~\cite{bs} has not 
analyzed a precise enough version of the originally intended QEK model. If I am right, the 
problem of in-principle validity of QEK remains open. Numerical tests might
discover a new candidate problem with QEK in the future which could invalidate the
quenching approach in principle. 

\section{Final comments.}

In this paper algorithmic issue in the QEK case have not been addressed.
Clearly, the $U(2)$ example was presented with traditional $SU(2)$ Monte Carlo 
updates in mind. An HMC version might be also worth looking into. 

From the extensive and ultimately successful work on the twisted EK model~\cite{tek}, TEK, it is known that for TEK to work, the large $N$ limit needs to be approached with care and
one needs to go to truly large values of $N$. By the law of 
``conservation of difficulty'' QEK may also need further nontrivial refinements.
The problem of annealed permutations BS~\cite{bs} found, at least at the theoretical level, 
seems harmless to me because the basic rules of calculus would tell 
you to eliminate permutations in the quenching approach and how to do it.

\begin{acknowledgments}
	I am grateful to Barak Bringolz for providing me, many years ago, with a succinct  
	formulation of the main point of the BS paper~\cite{bs}, 
	namely, that eq. (10) in~\cite{qek} showed that there was an issue 
	with permutations that could jeopardize the validity of QEK.  
	I thank Rajamani Narayanan for valuable discussions and comments.
\end{acknowledgments}

\section{References Cited}

\clearpage

\end{document}